\documentclass{article}
\pdfoutput=1
\PassOptionsToPackage{numbers, compress}{natbib}

\usepackage[preprint]{neurips_2024}

\usepackage[utf8]{inputenc} 
\usepackage[T1]{fontenc}  
\usepackage{xcolor}     
\usepackage{color}
\definecolor{ggreen}{HTML}{58a55c}
\usepackage[pagebackref=true,
   breaklinks=true,
   letterpaper=true,
   colorlinks = true,
   linkcolor = red,
   urlcolor = blue,
   citecolor = ggreen,
   anchorcolor = blue]{hyperref}
\usepackage{url}      
\usepackage{booktabs}    
\usepackage{amsfonts}    
\usepackage{nicefrac}    
\usepackage{microtype}   

\usepackage{algorithmic}
\usepackage{algorithm}

\usepackage{amsmath}
\usepackage{graphicx}
\usepackage{algorithm}
\usepackage{subcaption}
\usepackage{xcolor}
\usepackage{colortbl}

\usepackage{pifont}
\usepackage{tabu}
\usepackage{tabularx}
\definecolor{Gray}{gray}{0.9}
\definecolor{Cyan}{rgb}{0.88,1,1}
\usepackage{listings}
\usepackage{bm}
\usepackage{adjustbox}
\usepackage{ragged2e}
\usepackage{multirow}
\usepackage{amssymb}
\usepackage{wrapfig}
\usepackage{makecell}

\usepackage{soul}
\sethlcolor{Gray}

\definecolor{00red}{RGB}{236,35,35}

\definecolor{00blue}{RGB}{50,149,237}

\definecolor{00pink}{RGB}{200,151,225}
\definecolor{02pink}{RGB}{200,151,227}

\definecolor{00grey}{RGB}{166,166,166}

\definecolor{00green}{RGB}{82,208,83}
\definecolor{02green}{RGB}{83,209,86}

\definecolor{00pink}{RGB}{230,114,138}

\definecolor{00purple}{RGB}{219,103,219}

\numberwithin{equation}{section}

\newcolumntype{x}[1]{>{\centering\arraybackslash}p{#1pt}}
\newlength\savewidth

\newcommand{\PreserveBackslash}[1]{\let\temp=\\#1\let\\=\temp}
\newcolumntype{C}[1]{>{\PreserveBackslash\centering}p{#1}}
\newcolumntype{L}[1]{>{\PreserveBackslash\raggedright}p{#1}}

\usepackage{etoolbox}
\makeatletter
\AfterEndEnvironment{algorithm}{\let\@algcomment\relax}
\AtEndEnvironment{algorithm}{\kern2pt\hrule\relax\vskip3pt\@algcomment}
\let\@algcomment\relax
\newcommand\algcomment[1]{\def\@algcomment{\footnotesize#1}}
\renewcommand\fs@ruled{\def\@fs@cfont{\bfseries}\let\@fs@capt\floatc@ruled
 \def\@fs@pre{\hrule height.8pt depth0pt \kern2pt}%
 \def\@fs@post{}%
 \def\@fs@mid{\kern2pt\hrule\kern2pt}%
 \let\@fs@iftopcapt\iftrue}
\makeatother

\newcommand*\samethanks[1][\value{footnote}]{\footnotemark[#1]}

\newcommand\cb[1]{\color{blue} #1}

\title{UltraLight VM-UNet: Parallel Vision Mamba Significantly Reduces Parameters for Skin Lesion Segmentation}

\author{
Renkai Wu$^{1}$
~~
Yinghao Liu$^{2}$ 
~~
Pengchen Liang$^{1}$\thanks{Corresponding authors.}
~~
Qing Chang$^{1}$\samethanks
\\
$^{1}$Shanghai University ~~ $^{2}$University of Shanghai for Science and Technology
}

\begin{document}

\maketitle

\begin{abstract}
Traditionally for improving the segmentation performance of models, most approaches prefer to use adding more complex modules. And this is not suitable for the medical field, especially for mobile medical devices, where computationally loaded models are not suitable for real clinical environments due to computational resource constraints. Recently, state-space models (SSMs), represented by Mamba, have become a strong competitor to traditional CNNs and Transformers. In this paper, we deeply explore the key elements of parameter influence in Mamba and propose an UltraLight Vision Mamba UNet (UltraLight VM-UNet) based on this. Specifically, we propose a method for processing features in parallel Vision Mamba, named PVM Layer, which achieves excellent performance with the lowest computational complexity while keeping the overall number of processing channels constant. We conducted comparisons and ablation experiments with several state-of-the-art lightweight models on three skin lesion public datasets and demonstrated that the UltraLight VM-UNet exhibits the same strong performance competitiveness with parameters of only 0.049M and GFLOPs of 0.060. In addition, this study deeply explores the key elements of parameter influence in Mamba, which will lay a theoretical foundation for Mamba to possibly become a new mainstream module for lightweighting in the future. The code is available from~\url{https://github.com/wurenkai/UltraLight-VM-UNet}.
\end{abstract}

\section{Introduction}
With the continuous development of computer technology and hardware computing power, computer-aided diagnosis has been widely used in the medical field, and medical image segmentation is an important part of it. Medical image segmentation is usually realized using deep learning networks represented by convolution and Transformers. Convolution has excellent localized feature extraction capabilities, but is deficient in establishing the correlation of remote information \cite{rao2022hornet, rao2021global, wu2024mhorunet}. In the previous work \cite{liu2022convnet, ding2022scaling}, researchers proposed to utilize large convolutional kernels to alleviate this problem. As for the network architecture based on Transformers, in recent years, researchers \cite{hatamizadeh2022unetr, hatamizadeh2021swin} have deeply investigated its methods. Although the self-attention mechanism can solve the problem of remote information extraction by means of sequences of consecutive patches, it also introduces more computational complexity. This is because the quadratic complexity of the self-attention mechanism is closely related to the image size.

In addition, in terms of improving the accuracy of computer-aided diagnosis, the number of parameters of the algorithmic model is often enriched to improve the predictive power of the model \cite{aghdam2023attention, wu2024hsh}. However, for clinical and real healthcare environments, realistic computational power and memory constraints often need to be considered. Low parameters and minimal computational memory footprint are essential considerations in mHealth tasks \cite{ruan2023ege}. Therefore, an algorithmic model with low computational complexity and good performance is urgently needed for future mobile medical devices.

\begin{figure}[!t]
\centering
\includegraphics[width=\linewidth]{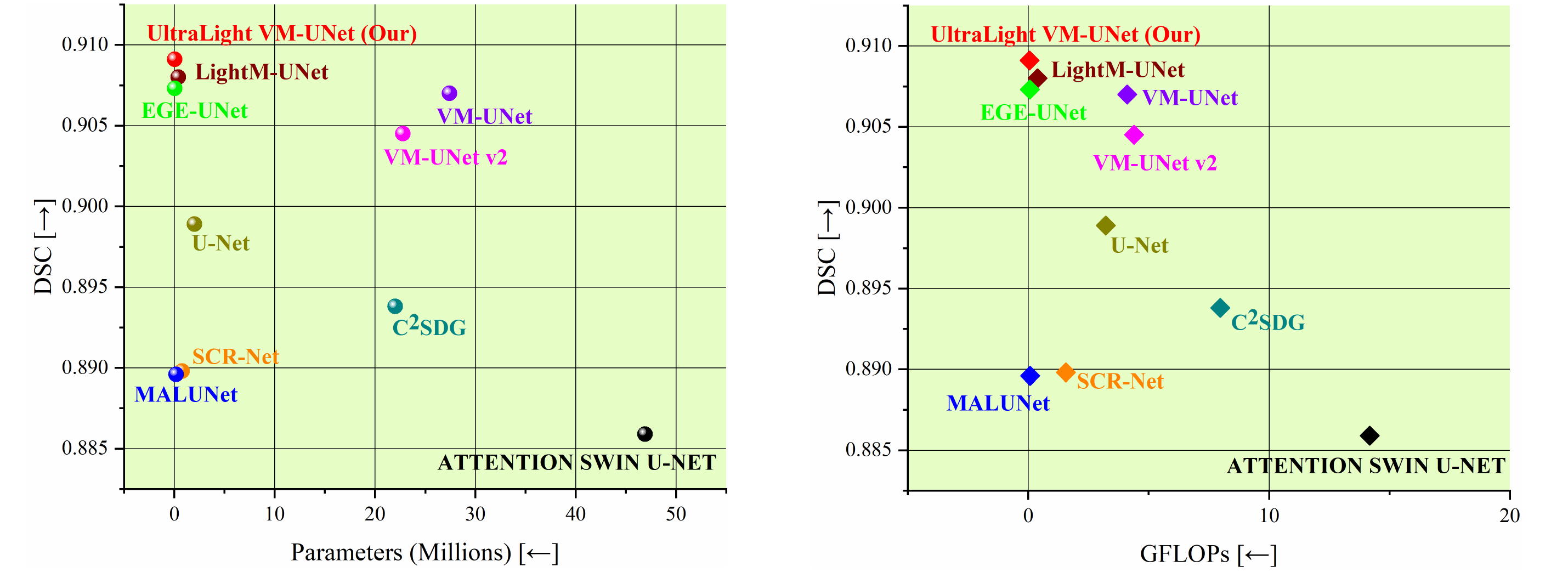}
\caption{Visualization of the comparison results for the ISIC2017 dataset. X-axis corresponds to parameters and GFLOPs, the fewer the better. Y-axis corresponds to segmentation performance (DSC), the higher the better.}
\label{fig01}
\end{figure}

Recently, state-space models (SSMs) have shown linear complexity in terms of input size and memory occupation \cite{liu2024vmamba,zhu2024vision}, which makes them key to lightweight model foundations. In addition, SSMs excel at capturing remote dependencies, which can critically address the problem of convolution for extracting information over long distances. In Gu et al. \cite{gu2023mamba}, time-varying parameters were introduced into SSM to obtain Mamba, and it was demonstrated that Mamba is able to process textual information with lower parameters than Transformers. On the vision side, the introduction of Vision Mamba \cite{zhu2024vision} has once again furthered people's understanding of Mamba, which saves 86.8$\%$ of memory when reasoning about images of 1248$\times$1248 size without the need for an attentional mechanism. With the outstanding work of the researchers mentioned above, we are more confident that Mamba will occupy a major position in the future as a basic building block for lightweight models.

\begin{figure}[!t]
\centering
\includegraphics[width=\linewidth]{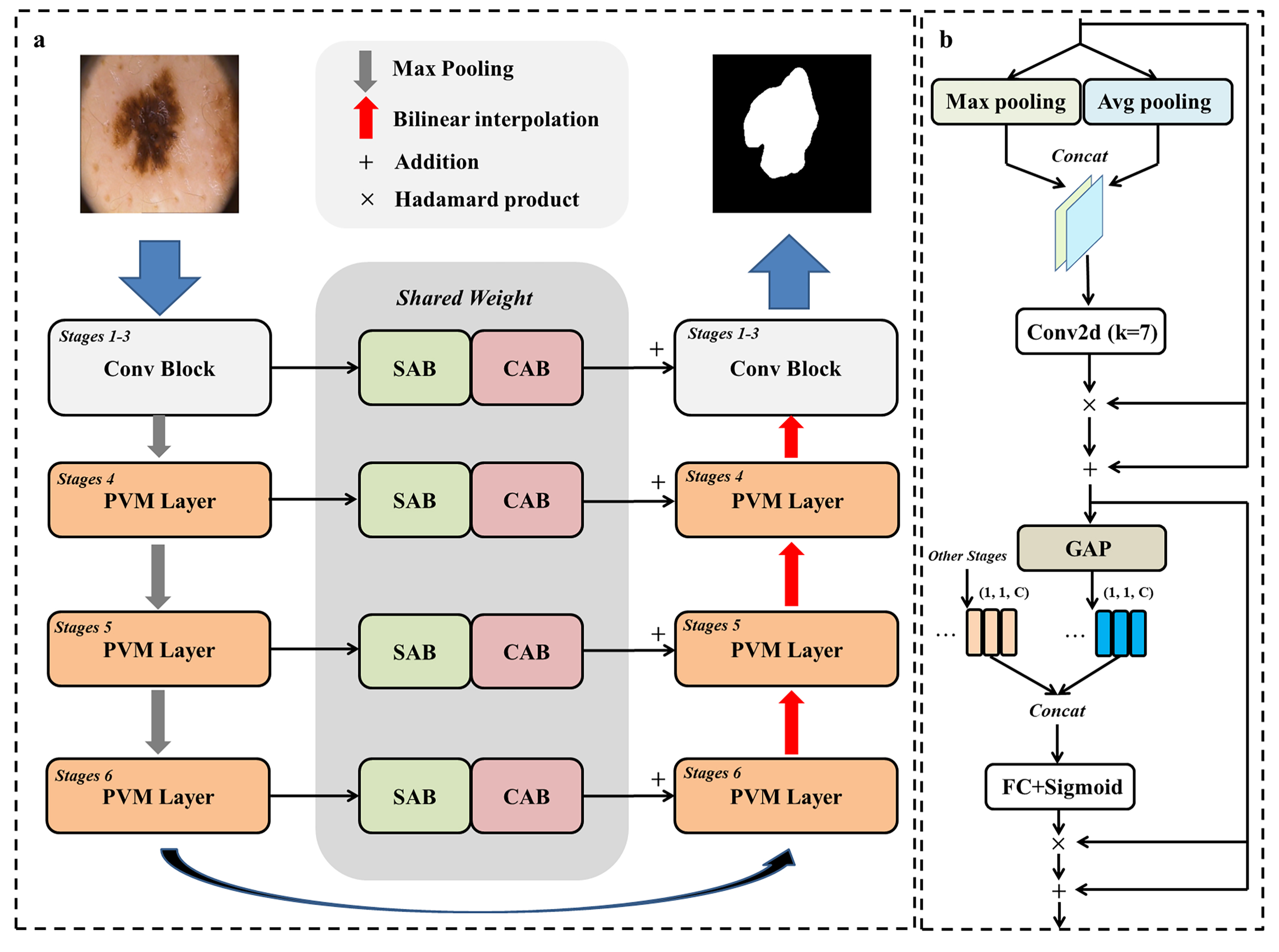}
\caption{(a) The proposed UltraLight Vision Mamba UNet (UltraLight VM-UNet) model architecture. (b) Multilevel and multiscale information fusion module architecture for skip-connection paths. }
\label{fig02}
\end{figure}

In this paper, we are building a lightweight model based on Vision Mamba. We deeply explore the critical memory footprint of Mamba and the performance tradeoffs, and propose an UltraLight Vision Mamba UNet (UltraLight VM-UNet). To the best of our knowledge, the proposed UltraLight VM-UNet is the most lightweight Vision Mamba model available (with parameters of 0.049M and GFLOPs of 0.060) and exhibits very competitive performance in three skin lesion segmentation tasks. Specifically, we delve into the keys affecting the computational complexity in Mamba, and conclude that the number of channels is a key factor in the explosive memory occupation for Mamba computation. We build on this finding of ours by proposing a parallel Vision Mamba approach for processing deep features, named PVM Layer, which simultaneously keeps the overall processing channel count constant. The proposed PVM Layer achieves excellent performance with surprisingly low parameters. In addition, the deep feature extraction of the proposed UltraLight VM-UNet we implement using only the PVM Layer containing Mamba, as shown in Figure \ref{fig02}. In the Methods section, we will present the details of the proposed UltraLight VM-UNet as well as the key factors of the parameter effects in Mamba and the performance balancing approach.

Although similar parallel connection of modules has been mentioned in previous studies \cite{szegedy2015going,ioffe2015batch,szegedy2016rethinking,szegedy2017inception}, the impact of using parallel connection in Mamba is still unknown. Does parallel connection lead to a significant performance degradation of Mamba when utilizing the SSM selection mechanism? This is because parallel connections lead to a reduction in the number of feature channels learned per SSM. In this paper, we will give the answer in detail. Parallel Vision Mamba or Mamba not only remains competitive in terms of performance, but gets a significant reduction in parameters and computational complexity.

Our contributions and findings can be summarized as follows:

\begin{itemize}
\item An UltraLight Vision Mamba UNet (UltraLight VM-UNet) is proposed for skin lesion segmentation. To the best of our knowledge, the UltraLight VM-UNet is the lightest Vision Mamba model available (parameters of 0.049M, GFLOPs of 0.060).

\item A parallel Vision Mamba method for processing deep features, named PVM Layer, is proposed, which achieves excellent performance with the lowest computational complexity while keeping the overall number of processing channels constant. This can be generalized to the parallel connection of any Mamba variant.

\item We provide an in-depth analysis of the key factors influencing the parameters of Mamba, and provide a theoretical basis for Mamba to become a mainstream module for lightweight modeling in the future. 

\item The proposed UltraLight VM-UNet parameters are 99.82$\%$ lower than the traditional pure Vision Mamba UNet model (VM-UNet) and 87.84$\%$ lower than the parameters of the current lightest Vision Mamba UNet model (LightM-UNet). In addition, the UltraLight VM-UNet maintains strong performance competitiveness in all three publicly available skin lesion segmentation datasets.

\end{itemize}

\section{Related Work}
With the significant improvement of computer computing power, computer vision has become an important field in computer technology nowadays. And deep learning development has led to the full convolutional model (FCN) \cite{long2015fully} showing excellent performance in image segmentation methods. Soon after the emergence of FCN, another fully convolutional model (U-Net) \cite{ronneberger2015u} emerged to generate renewed excitement. The skip-connection operation in U-Net is able to merge high-level and low-level features well, which is especially important for image segmentation, especially for medical image segmentation that requires fine-grained segmentation.

Medical image segmentation, as one of the important branches in image segmentation, is also one of the research directions to which many researchers have devoted their efforts. Among them, multi-scale variation problem and feature refinement learning are the key problems in medical image segmentation \cite{xiao2023transformers}. And skin lesion segmentation has rich and varied feature information as well as high lethality caused by its malignant melanoma \cite{siegel2022cancer}, which has led many researchers to carry out a series of studies around skin lesion segmentation \cite{aghdam2023attention,wu2024mhorunet,wu2024hsh}.

Medical image segmentation algorithms represented by skin lesions have been rapidly developed after the advent of U-Net. In Aghdam et al. \cite{aghdam2023attention}, an inhibition operation of the attention mechanism in cascade operation has been proposed for skin lesion segmentation based on Swin U-Net \cite{cao2022swin}. MHorUNet \cite{wu2024mhorunet} model proposes a high-order spatial interaction UNet model for skin lesion segmentation. In Wu et al. \cite{wu2024hsh}, an adaptive selection of higher order UNet model for order interaction has been proposed for skin lesion segmentation. In addition, there are very many algorithms based on U-Net improved for skin lesion segmentation. However, it is common for researchers to add richer modules to the model to improve the accuracy of recognition, but this also significantly increases the parameters and computational complexity of the model. After the emergence of Vision Mamba, LightM-UNet \cite{liao2024lightm} was proposed to reduce the number of parameters in the model based on Mamba. LightM-UNet further extracts the deep semantics and tele-relationships by using the residual Vision Mamba, and achieves better performance with a smaller number of parameters. In addition, U-Mamba \cite{ma2024u} introduced Vision Mamba into the U-framework for the first time, but its having a large number of parameters (173.53M) limits its use in real clinical settings. 

In this paper, in order to solve the current problem of large model parameters and to reveal the key factors affecting Mamba parameters. We propose an UltraLight Vision Mamba UNet (UltraLight VM-UNet) based on Mamba with a parameter of only 0.049M. The UltraLight VM-UNet is confirmed to still maintain a strong competitive performance in three public datasets of skin lesions. In the next section, our method is described in detail.

\section{Method}
\subsection{Architecture Overview}
The proposed UltraLight Vision Mamba UNet (UltraLight VM-UNet) is shown in Figure \ref{fig02}. The UltraLight VM-UNet has a total of 6-layer structure consisting of a U-shaped structure (encoder, decoder, and skip-connection path). The number of channels in the 6-layer structure is set to [8, 16, 24, 32, 48, 64]. The extraction of shallow features in the first 3 layers is composed using a convolution module (Conv Block), where each layer includes a standard convolution with a convolution kernel of 3 and a maximum pooling operation. The deeper features from layer 4 to layer 6 are our core part, where each layer consists of our proposed Parallel Vision Mamba Layer (PVM Layer). The decoder part maintains the same setup as the encoder. The skip-connection path utilize the Channel Attention Bridge (CAB) module and the Spatial Attention Bridge (SAB) module for multilevel and multiscale information fusion \cite{ruan2022malunet}.

\subsection{Mamba Parameter Impact Analysis}
\begin{figure}[!t]
\centering
\includegraphics[width=\linewidth]{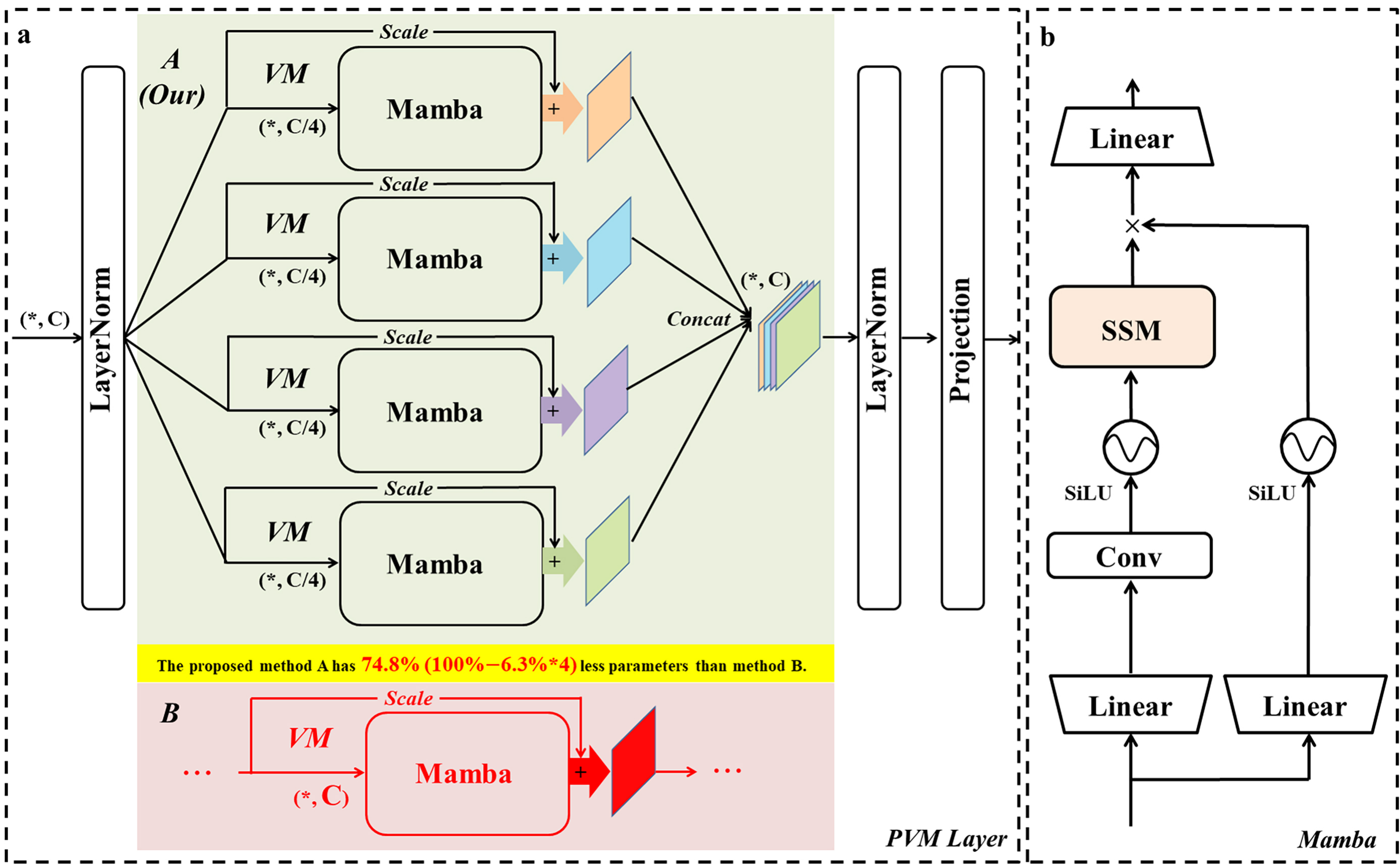}
\caption{(a) Architecture of the proposed Parallel Vision Mamba Layer (PVM Layer) method. Vision Mamba (VM) is composed by Mamba combined with residual connection and adjustment factor. (b) Mamba composition structure. }
\label{fig03}
\end{figure}

Vision Mamba for PVM Layer is mainly composed using Mamba combined with residual connections and adjustment factors (Figure \ref{fig03}), which allows traditional Mamba to improve the capture of remote spatial relations without introducing additional parameters and computational complexity \cite{liao2024lightm}. This has a better improvement in the performance of Mamba in visual tasks while keeping the parameter and computational complexity low.

Among SSM-based Mamba, the number of channels, the size of the SSM state dimension, the size of the internal 1D convolutional kernel, the projection dilation multiplier, and the rank of the step size all affect the parameters. And in this, the impact of the channel number is explosive, and its main influence is from the following multiple directions:

First, the $d\_inner$ of the Mamba internal extended projection channel is determined by the product of the projection expansion multiplier and the number of input channels. This can be specifically expressed by the following equation:
\begin{equation}
d\_inner = expand *  d\_model 
\end{equation}
\noindent
where $d\_inner$ is the internal expansion projection channel, $expand$ is the projection expansion multiplier (fixed at 2 by default), and $d\_model$ is the number of input channels. We can conclude that $d\_inner$ will double up as the number of channels ($d\_model$) per layer in the model increases.

Second, the parameters of the input projection layer (the same input linear layer is used for both branches) and output projection layer within Mamba will be directly related to the number of input channels. The input projection layer and output projection layer operate as follows:
\begin{equation}
in\_proj : nn.Linear(d\_model, d\_inner * 2, bias=False) 
\end{equation}
\begin{equation}
out\_proj : nn.Linear(d\_inner, d\_model, bias=False)
\end{equation}
\noindent
where then the input projection ($in\_proj$) layer parameter is $d\_model * d\_inner * 2$, and the output projection layer ($out\_proj$) parameter is $d\_inner * d\_model$. we can conclude that the number of input channels, $d\_model$, is the key element controlling the parameter, where the internal extended projection channel, $d\_inner$, is also controlled by $d\_model$ is controlled.

Further, the intermediate linear projection layers of SSM are also key to the influence of the parameters. The details are as follows:
\begin{equation}
x\_proj = nn.Linear(d\_inner, (dt\_rank + d\_state * 2), bias=False) 
\end{equation}
\begin{equation}
dt\_proj = nn.Linear(dt\_rank, d\_inner, bias=True)
\end{equation}
\noindent
where $dt\_rank$ is the rank of the step $(dt\_rank = d\_model/16)$, $d\_state$ is the size of the state dimension (fixed to 16), where the parameters can be derived as $d\_inner * (dt\_rank + d\_state*2)$. $dt\_proj$ is a linear projection layer for step size, with parameters $(dt\_rank * d\_inner) + d\_inner$, mainly used for linear projection for step size ($dt$). So, we can conclude that all parameters are still mainly controlled by the number of input channels $d\_model$.

In addition, the internal convolution ($nn.Conv1d(d\_inner,d\_inner,d\_conv, bias=True$)) also provides parametric influence with $d\_inner * d\_inner * d\_conv + d\_inner$. In this paper, $d\_conv$ is fixed to 4, so the convolution provides a parameter of $4 * d\_inner^{2} + d\_inner$, which is also controlled by the $d\_model$.

Also, the logarithmic form parameters ($A\_logs$) of the transfer matrix $A$ in the SSM module are an important influencing element. $A\_logs$ is a parameter matrix of the shape ($d\_inner$, $d\_state$), so its parameters can be derived as $d\_inner * d\_state$. In addition, the trainable vector parameter ($D$) of the process computed within the SSM contains the $d\_inner$ parameter, which is used to selectively integrate the SSM state outputs with the original input signals, thereby enhancing the model's expressiveness and training stability.

In summary, assuming that the original channel number is 1024, keeping other parameters unchanged, and when the channel number is reduced to one-fourth of the original (channel number 256), the original total parameters can be calculated from the above parameter formula to get the original total parameters reduced from 23,435,264 to 1,484,288. The parameter explosion is reduced by 93.7$\%$, which further confirms that the number of channels has a very critical impact on Mamba parameters.

Based on the above in-depth study of the key elements affecting the Mamba parameters, we propose a method for processing features in parallel Vision Mamba, named PVM Layer. Excellent performance is achieved with the lowest computational complexity, while keeping the overall number of processing channels unchanged. Specifically, it will be detailed in the next section.

\subsection{Parallel Vision Mamba Layer}
As analyzed in the previous subsection, the number of input channels has an explosive effect on the parameters of Mamba. As shown in Figure \ref{fig03}(a), we propose the Parallel Vision Mamba Layer (PVM Layer) for processing deep features. Specifically, a feature $X$ with channel number $C$ first passes through a LayeNorm layer and then is divided into $Y_{1}^{C / 4}$, $Y_{2}^{C / 4}$, $Y_{3}^{C / 4}$ and $Y_{4}^{C / 4}$ features each with channel number $C/4$. After that, each of the features is then fed into Mamba, and then the output is subjected to residual concatenation and adjustment factor for optimizing the remote spatial information acquisition capability \cite{liao2024lightm}. Finally, the four features are combined into the feature $X_{out}$ with channel number $C$ by concat operation, and then output by LayerNorm and Projection operation respectively. The specific operation can be expressed by the following equation:
\begin{equation}
Y_{1}^{C / 4}, Y_{2}^{C / 4}, Y_{3}^{C / 4}, Y_{4}^{C / 4}=Sp\left[L N\left(X_{in}^{C}\right)\right]
\end{equation}
\begin{equation}
VM\_ Y_{i}^{C / 4}=Mamba\left(Y_{i}^{C / 4}\right)+\theta \cdot Y_{i}^{C / 4} \quad i=1,2,3,4
\end{equation}
\begin{equation}
X_{out }=Cat\left(VM_{-} Y_{1}^{C/4}, VM_{-} Y_{2}^{C / 4}, VM_{-} Y_{3}^{C / 4}, VM_{-} Y_{4}^{C / 4}\right)
\end{equation}
\begin{equation}
Ou t=Pro\left[L N\left(X_{out }\right)\right]
\end{equation}
\noindent
where $LN$ is the LayerNorm, $Sp$ is the Split operation, $Mamba$ is the Mamba operation, $\theta$ is the adjustment factor for the residual connection, $Cat$ is the concat operation, and $Pro$ is the Projection operation. From Eq. 3.7, we used parallel Vision Mamba processing features, while ensuring that the total number of channels processed remains constant, maintaining high accuracy while maximizing parameter reduction. As shown in Figure \ref{fig03}(a) for Methods A and B, again assuming a channel count size of 1024, each Vision Mamba (VM) in Method A reduces the parameters by 93.7$\%$. And it contains 4 such operations, so when summed up, the comparison method B parameters are reduced by 74.8$\%$ overall. Through our proposed parallel Vision Mamba operation, the parameter reduction is maximized while maintaining strong performance competitiveness.

\subsection{Skip-connection Path}
The skip-connection path uses the Spatial Attention Bridge (SAB) module and the Channel Attention Bridge (CAB) module proposed by Ruan et al. \cite{ruan2022malunet}, as shown in Figure \ref{fig02}(b). The combined use of SAB and CAB allows for the fusion of multi-stage features of different scales of the UltraLight VM-UNet. The SAB module contains the maximum pooling, the average pooling, the extended convolution of the shared weights. The CAB module contains global average pooling, concat operation, fully connected layers, and sigmoid activation function. Both SAB and CAB have been shown to be effective in improving the convergence ability of the model and enhancing the sensitivity to lesions in previous work \cite{ruan2022malunet, wu2024mhorunet, wu2024h}.

\section{Experiment}
\subsection{Comparison results}
In order to validate the competitive performance of the proposed UltraLight VM-UNet under the 0.049M parameter, we conducted comparison experiments with several state-of-the-art lightweight and classical medical image segmentation models. Specifically, they include U-Net \cite{ronneberger2015u}, SCR-Net \cite{wu2021precise}, ATTENTION SWIN U-NET \cite{aghdam2023attention}, C$^2$SDG \cite{hu2023devil}, VM-UNet \cite{ruan2024vm}, VM-UNet v2 \cite{zhang2024vm}, MALUNet \cite{ruan2022malunet}, LightM-UNet \cite{liao2024lightm} and EGE-UNet \cite{ruan2023ege}.

\begin{table}[t]
 \centering
 \caption{\small \textbf{Experimental comparisons of the proposed UltraLight VM-UNet with the best lightweight and classical models.} This includes comparisons on two publicly available large dermatologic lesion datasets (ISIC2017 and ISIC2018), as well as external validation on a small publicly available dataset (PH$^2$). “\emph{↑}” stands for external validation experiments. In addition, comparisons of the parameters and computational complexity of the individual models are included. Our models are highlighted in \hl{gray}. }

\newcommand{\midsepnew}{\aboverulesep = 0.605mm \belowrulesep = 0.984mm}
\newcommand{\midsepdefault}{\aboverulesep = 0.605mm \belowrulesep = 0.984mm}
 \begin{minipage}{.49\textwidth}
 \adjustbox{width=\linewidth}{
 \midsepnew
 \setlength{\tabcolsep}{2pt}
  \begin{tabular}{L{120pt}L{35pt}L{30pt}L{30pt}L{40pt}}\toprule
  \multicolumn{5}{l}{\emph{Comparison experiments on the ISIC2017 dataset.}} \\
  \midrule
  Model & DSC & SE & SP & ACC \\ 
  \midrule
  U-Net     & 0.8989          & 0.8793          & 0.9812          & 0.9613 \\
  SCR-Net      & 0.8898          & 0.8497          & \textbf{0.9853} & 0.9588  \\
  ATTENTION SWIN U-NET   & 0.8859          & 0.8492          & 0.9847          & 0.9591 \\
  C$^2$SDG    & 0.8938          & 0.8859          & 0.9765          & 0.9588  \\
  VM-UNet      & 0.9070           & 0.8837          & 0.9842          & 0.9645 \\
  VM-UNet v2    & 0.9045          & 0.8768          & 0.9849          & 0.9637  \\
  MALUNet   & 0.8896          & 0.8824          & 0.9762          & 0.9583 \\
  LightM-UNet   & 0.9080           & 0.8839          & 0.9846          & \textbf{0.9649} \\ 
  EGE-UNet    & 0.9073          & 0.8931          & 0.9816          & 0.9642  \\
  \rowcolor{Gray}UltraLight VM-UNet (Our) & \textbf{0.9091} & \textbf{0.9053} & 0.9790           & 0.9646  \\
  \midrule
  \midrule
  \multicolumn{5}{l}{\emph{Comparison experiments on the ISIC2018 dataset.}} \\
  \midrule
  Model & DSC & SE & SP & ACC \\ 
  \midrule
  U-Net     & 0.8851                           & 0.8735                          & 0.9744                          & 0.9539 \\
  SCR-Net      & 0.8886                           & 0.8892                          & 0.9714                      & 0.9547  \\
  ATTENTION SWIN U-NET   & 0.8540                     & 0.8057                     & \textbf{0.9826}              & 0.9480 \\
  C$^2$SDG    & 0.8806                  & 0.8970                 & 0.9643                          & 0.9506  \\
  VM-UNet      & 0.8891       & 0.8809                        & 0.9743                          & 0.9554 \\
  VM-UNet v2    & 0.8902          & 0.8959          & 0.9702          & 0.9551  \\
  MALUNet   & 0.8931             & 0.8890                          & 0.9725                          & 0.9548 \\
  LightM-UNet   & 0.8898                         & 0.8829           & 0.9765                          & 0.9555 \\ 
  EGE-UNet    & 0.8819          & \textbf{0.9009}          & 0.9638          & 0.9510  \\
  \rowcolor{Gray}UltraLight VM-UNet (Our) & \textbf{0.8940} & 0.8680 & 0.9781           & \textbf{0.9558}  \\
  \bottomrule
   \end{tabular}%
   \midsepdefault
    }
  \end{minipage}%
  \hfill
  \begin{minipage}{.49\textwidth}
 \adjustbox{width=\linewidth}{
 \midsepnew
 \setlength{\tabcolsep}{2pt}
  \begin{tabular}{L{120pt}L{35pt}L{30pt}L{30pt}L{40pt}}\toprule
  \multicolumn{5}{l}{\emph{↑ Comparison experiments on the PH$^2$ dataset.}} \\
  \midrule
  Model & DSC & SE & SP & ACC \\ 
  \midrule
  U-Net     & 0.9060                           & 0.9255                          & 0.9440                          & 0.9381 \\
  SCR-Net     & 0.8989                           & 0.9114                   & 0.9446                          & 0.9339  \\
  ATTENTION SWIN U-NET   & 0.8850                  & 0.8886                & 0.9363                          & 0.9213 \\
  C$^2$SDG    & 0.9030                           & 0.9137          & 0.9476                          & 0.9367  \\
  VM-UNet      & 0.9033           & 0.9131                          & 0.9483                          & 0.9369 \\
  VM-UNet v2    & 0.9050          & 0.9160          & 0.9485          & 0.9380  \\
  MALUNet   & 0.8865            & 0.8922                          & 0.9425                          & 0.9263 \\
  LightM-UNet   & 0.9156                           & 0.9129    & 0.9613                          & 0.9457 \\ 
  EGE-UNet    & 0.9086          & 0.9198          & 0.9502          & 0.9404  \\
  \rowcolor{Gray}UltraLight VM-UNet (Our) & \textbf{0.9265} & \textbf{0.9345} & \textbf{0.9606}           & \textbf{0.9521}  \\
  \midrule
  \midrule
  \multicolumn{5}{l}{\emph{Comparison of parameters and computational consumption.}} \\
  \midrule
  Model & \multicolumn{2}{c}{Params} & \multicolumn{2}{c}{GFLOPs}  \\ 
  \midrule
  U-Net     & \multicolumn{2}{c}{2.009M} & \multicolumn{2}{c}{3.224} \\
  SCR-Net     & \multicolumn{2}{c}{0.801M} & \multicolumn{2}{c}{1.567}  \\
  ATTENTION SWIN U-NET   & \multicolumn{2}{c}{46.910M} & \multicolumn{2}{c}{14.181} \\
  C$^2$SDG    & \multicolumn{2}{c}{22.001M} & \multicolumn{2}{c}{7.972}  \\
  VM-UNet     & \multicolumn{2}{c}{27.427M} & \multicolumn{2}{c}{4.112} \\
  VM-UNet v2    & \multicolumn{2}{c}{22.771M} & \multicolumn{2}{c}{4.400}  \\
  MALUNet   & \multicolumn{2}{c}{0.175M} & \multicolumn{2}{c}{0.083} \\
  LightM-UNet   & \multicolumn{2}{c}{0.403M} & \multicolumn{2}{c}{0.391} \\ 
  EGE-UNet    & \multicolumn{2}{c}{0.053M} & \multicolumn{2}{c}{0.072}  \\
  \rowcolor{Gray}UltraLight VM-UNet (Our) & \multicolumn{2}{c}{\textbf{0.049M}} & \multicolumn{2}{c}{\textbf{0.060}}  \\
  \bottomrule
   \end{tabular}%
   \midsepdefault
  }
  \end{minipage}
  
 \label{tab1}%
  \vspace{-10pt}
\end{table}%

\begin{figure}[!t]
\centering
\includegraphics[width=\linewidth]{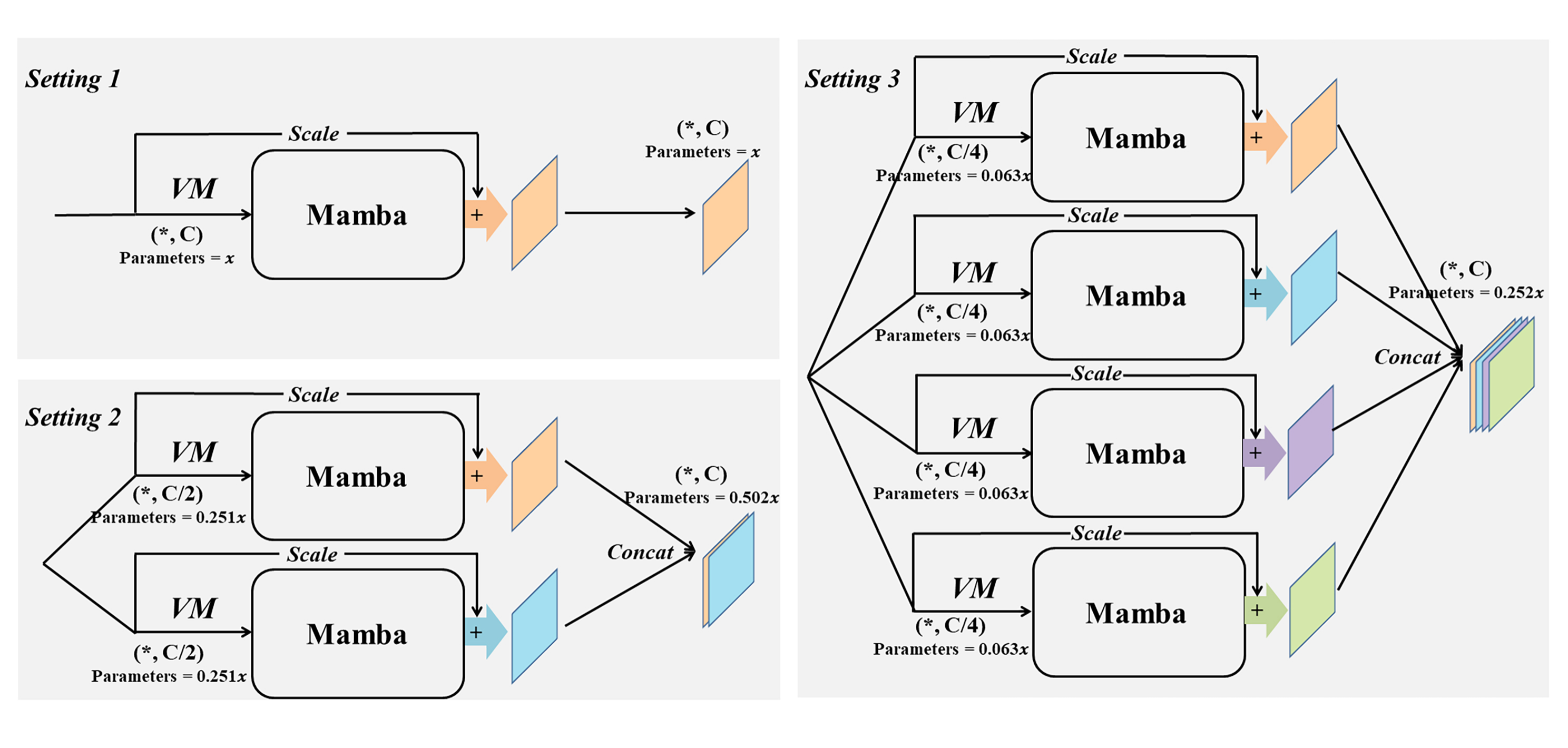}
\caption{Settings for ablation experiments with Vision Mamba used in different parallel ways (PVM Layer). }
\label{fig04}
\end{figure}

Table \ref{tab1} show the experimental results on ISIC2017, ISIC2018 and PH$^2$ datasets, respectively. As shown in the table, the parameters of our model are 99.82$\%$ lower than those of the traditional pure Vision Mamba UNet model (VM-UNet) and 87.84$\%$ lower than those of the current lightest Vision Mamba UNet model (LightM-UNet). In addition, the GFLOPs of our model are 98.54$\%$ lower than VM-UNet and 84.65$\%$ lower than LightM-UNet. With such a large reduction in parameters and GFLOPs, the performance of our model still maintains excellent and highly competitive performance. In addition, MALUNet is a lightweight model proposed based on convolution, and although it has lower parameters and GFOLPs than VM-UNet and LightM-UNet, the parameters and GFOLPs of our model are still 72.0$\%$ and 27.71$\%$ lower than them, respectively. In particular, the performance of both MALUNet, the proposed lightweight model based on convolution, is much lower than that of the Mamba-based model, which reflects that it is difficult for the convolution-based lightweight model to balance the relationship between performance and computational complexity. 

\begin{table}[htbp]
  \centering
  \caption{Ablation experiments on the effect of Vision Mamba in different parallel connections.}
    \begin{tabu}to 1.0\textwidth{*{7}{X[c]}}\toprule
    Settings  & Params & GFLOPs & DSC & SE & SP & ACC \\\midrule
    1                 & 0.136M                          & 0.060           & 0.9069          & 0.8861          & 0.9834          & 0.9644          \\
2                 & 0.070M                          & 0.060           & 0.9073          & 0.8866          & \textbf{0.9835} & 0.9645          \\
3                 & \textbf{0.049M}                 & \textbf{0.060}  & \textbf{0.9091} & \textbf{0.9053} & 0.9790          & \textbf{0.9646} \\\bottomrule
    \end{tabu}%
    \vspace{-8pt}
  \label{tab2}%
\end{table}%

\begin{table*}[htbp]
\centering
\caption{\textbf{Impact of adopting parallelism for different SSM variants.} (\emph{P}) shows the adoption of quadruple parallelism to replace the original form. {\color{02green}$\blacksquare$} indicates experiments on the ISIC2017 dataset, {\color{00blue}$\spadesuit$} indicates the ISIC2018 dataset, and {\color{orange}$\clubsuit$} indicates the PH$^2$ dataset.}
\renewcommand{\arraystretch}{1}
\tiny
\adjustbox{width=\textwidth}{
\begin{tabular}{c|c|c|c|c|c|c|c|c}
\toprule
Methods & Parallelism & SSM-Variants  & Params & GFLOPs & DSC & SE & SP & ACC \\\midrule
{\color{02green}$\blacksquare$} VM-UNet & False & SS2D  & 27.427M & 4.112 & 0.9070 & 0.8837 & 0.9842 & 0.9645 \\
{\color{00blue}$\spadesuit$} VM-UNet& False & SS2D  & 27.427M & 4.112  & 0.8891 & 0.8809 & 0.9743 & 0.9554 \\
{\color{orange}$\clubsuit$} VM-UNet& False & SS2D  & 27.427M & 4.112  & 0.9033 & 0.9131 & 0.9483 & 0.9369 \\\midrule
\rowcolor{Gray}{\color{02green}$\blacksquare$} VM-UNet& True & (\emph{P}) SS2D  & 5.116M & 1.564 & 0.9159 & 0.8843 & 0.9886 & 0.9682 \\
\rowcolor{Gray}{\color{00blue}$\spadesuit$} VM-UNet& True & (\emph{P}) SS2D  & 5.116M & 1.564  & 0.8977 & 0.8996 & 0.9734 & 0.9584 \\
\rowcolor{Gray}{\color{orange}$\clubsuit$} VM-UNet& True & (\emph{P}) SS2D  & 5.116M & 1.564  & 0.9185 & 0.9340 & 0.9525 & 0.9465 \\
\midrule
\midrule
{\color{02green}$\blacksquare$} H-vmunet & False & H-SS2D  & 8.973M & 0.742 & 0.9172 & 0.9056 & 0.9831 & 0.9680 \\
{\color{00blue}$\spadesuit$} H-vmunet& False & H-SS2D  & 8.973M & 0.742  & 0.8966 & 0.8952 & 0.9741 & 0.9581 \\
{\color{orange}$\clubsuit$} H-vmunet& False & H-SS2D  & 8.973M & 0.742  & 0.9146 & 0.9295 & 0.9509 & 0.9440 \\\midrule
\rowcolor{Gray}{\color{02green}$\blacksquare$} H-vmunet& True & (\emph{P}) H-SS2D  & 1.764M & 0.318 & 0.9066 & 0.8825 & 0.9843 & 0.9644 \\
\rowcolor{Gray}{\color{00blue}$\spadesuit$} H-vmunet& True & (\emph{P}) H-SS2D  & 1.764M & 0.318  & 0.8948 & 0.9067 & 0.9695 & 0.9567 \\
\rowcolor{Gray}{\color{orange}$\clubsuit$} H-vmunet& True & (\emph{P}) H-SS2D  & 1.764M & 0.318  & 0.9181 & 0.9254 & 0.9569 & 0.9467 \\

\bottomrule
\end{tabular}
}
\vspace{-10pt}
\label{tab3}%
\end{table*}

\begin{table}[t]
 \centering 
 \caption{\textbf{Comparative experiments where PVM Layer directly replaces modules from different models,} with results including parameters, computational complexity, and performance. \emph{Italics} in the method column (e.g., \emph{\_Conv}) indicate that the PVM Layer replaces the convolution module. In particular, we replace modules containing Convolution, Vision Transformer, Mamba, and Vision Mamba, covering the four most commonly used base modules. D$^{\rm P}$ and D$^{\rm G}$ denote the percentage decrease in parameters and GFLOPs, respectively, after replacing with PVM Layer.}
 \adjustbox{width=\linewidth}{
  \begin{tabular}{lC{45pt}C{45pt}C{45pt}C{45pt}C{45pt}C{45pt}C{45pt}C{45pt}}
  \toprule 
  \multicolumn{1}{l}{\multirow{2}[3]{*}{Methods}} & \multicolumn{4}{c}{\emph{Parameters and Computational Complexity}}  & \multicolumn{4}{c}{\emph{Performance Evaluation}} \\
  \cmidrule(lr){2-5}\cmidrule(lr){6-9}
     & Params &  D$^{\rm P}$ & GFLOPs & D$^{\rm G}$ & DSC & SE & SP & ACC \\
  \midrule
  UNet~\cite{ronneberger2015u}~~~~~~~~~~~~~~~~~~~~~~~~~~ & 2.009M & \multirow{2}{*}{80.38$\%$} & 3.224  & \multirow{2}{*}{78.74$\%$} & 0.8989 & 0.8793 & 0.9812  & 0.9613 \\
  UNet\_\emph{Conv}~ & 0.394M &  & 0.686 &  & 0.8974 & 0.8667 & 0.9842  & 0.9612 \\\midrule
  Att UNet~\cite{oktay2018attention} & 3.581M  & \multirow{2}{*}{45.10$\%$}  & 8.575   &  \multirow{2}{*}{29.61$\%$}  & 0.8821 & 0.8423  & 0.9836 & 0.9560 \\
  Att UNet\_\emph{Conv\_block} & 1.966M &  & 6.036  &  & 0.8857  & 0.8414  & 0.9857 & 0.9575\\\midrule
  SCR-Net~\cite{wu2021precise} & 0.812M & \multirow{2}{*}{81.90$\%$} & 1.567  & \multirow{2}{*}{72.75$\%$} & 0.8898 & 0.8497 & 0.9853 & 0.9588 \\
  SCR-Net\_\emph{Conv} & 0.147M & & 0.427  &  & 0.9058 & 0.8847 & 0.9833 & 0.9640 \\\midrule
  Swin-UNet~\cite{cao2022swin} & 27.176M & \multirow{2}{*}{76.11$\%$} & 7.724  & \multirow{2}{*}{75.48$\%$} & 0.8670 & 0.8427 & 0.9754 & 0.9494 \\
  Swin-UNet\_\emph{Vision Transformer} & 6.491M & & 1.894  &  & 0.8734 & 0.8445 & 0.9783 & 0.9521 \\\midrule
  META-Unet~\cite{wu2023meta} & 22.209M & \multirow{2}{*}{94.77$\%$} & 5.14  & \multirow{2}{*}{91.91$\%$} & 0.9068 & 0.8801 & 0.9836 & 0.9639 \\
  META-Unet\_\emph{ResNet Layer} & 1.161M &  & 0.416  & & 0.9047 & 0.8915 & 0.9807 & 0.9633 \\\midrule
  MHorUNet~\cite{wu2024mhorunet} & 9.585M & \multirow{2}{*}{90.35$\%$} & 0.864 & \multirow{2}{*}{81.94$\%$} & 0.9132 & 0.8974 & 0.9834 & 0.9666 \\
  MHorUNet\_\emph{Horblock} & 0.925M & & 0.156 & & 0.9107 & 0.8806 & 0.9870 & 0.9662 \\\midrule
  HSH-UNet~\cite{wu2024hsh} & 18.803M & \multirow{2}{*}{84.64$\%$}  & 9.362 & \multirow{2}{*}{90.58$\%$} & 0.9108 & 0.8907 & 0.9864 & 0.9654 \\
  HSH-UNet\_\emph{HSHB} & 2.888M &  & 0.882 &  & 0.9001 & 0.8938 & 0.9776 & 0.9612 \\\midrule
  MALUNet~\cite{ruan2022malunet} & 0.175M & \multirow{2}{*}{26.86$\%$} & 0.083 & \multirow{2}{*}{10.84$\%$} & 0.8896 & 0.8824 & 0.9762 & 0.9583 \\
  MALUNet\_\emph{DGA} & 0.128M & & 0.074 &  & 0.9025 & 0.8623 & 0.9882 & 0.9635 \\\midrule
  EGE-UNet~\cite{ruan2023ege} & 0.053M & \multirow{2}{*}{1.89$\%$} & 0.072 & \multirow{2}{*}{1.39$\%$} & 0.9073 & 0.8931 & 0.9816 & 0.9642 \\
  EGE-UNet\_\emph{GHPA} & 0.052M & & 0.071 &  & 0.9092 & 0.8941 & 0.9823 & 0.9650 \\\midrule
  VM-UNet~\cite{ruan2024vm} & 27.427M & \multirow{2}{*}{60.94$\%$} & 4.112 & \multirow{2}{*}{56.1$\%$} & 0.9070 & 0.8837 & 0.9842 & 0.9645 \\
  VM-UNet\_\emph{Vision Mamba} & 10.713M & & 1.805 &  & 0.8885 & 0.8517 & 0.9841 & 0.9582 \\\midrule
  VM-UNet v2~\cite{zhang2024vm} & 22.771M & \multirow{2}{*}{73.69$\%$} & 4.400 & \multirow{2}{*}{65.43$\%$} & 0.9045 & 0.8768 & 0.9849 & 0.9637 \\
  VM-UNet v2\_\emph{Vision Mamba} & 5.991M & & 1.521 &  & 0.8916 & 0.8692 & 0.9804 & 0.9586 \\\midrule
  LightM-UNet~\cite{liao2024lightm} & 0.403M & \multirow{2}{*}{72.95$\%$} & 0.391 & \multirow{2}{*}{0.26$\%$} & 0.9080 & 0.8839 & 0.9846 & 0.9649 \\
  LightM-UNet\_\emph{Mamba Layer} & 0.109M & & 0.390 &  & 0.9045 & 0.8668 & 0.9878 & 0.9642 \\\bottomrule
  \end{tabular}%
  }
   \vspace{-10pt}

 \label{tab4}%
\end{table}%

\subsection{Ablation experiments}
\textbf{Vision Mamba with different levels of parallelism.} In order to verify the validity of the proposed method of Vision Mamba (VM) with different parallelism, we performed a series of ablation experiments. As shown in Figure \ref{fig04}, we performed 3 different Settings. Setting 1 is a conventional connection of VM, Setting 2 is a connection using parallel connection of two VM with half the number of channels, and Setting 3 is a connection using parallel connection of four VM each with $C/4$ the number of channels. By analyzing the parameters of Mamba in section 3.2 of this study, assuming that the parameter of this module is $x$ for Setting 1 of the traditional VM connection method, Setting 2 can be calculated with a parameter of $0.502x$ and Setting 3 with a parameter of $0.252x$. Table \ref{tab2} shows the results of this ablation experiment, and it should be noted that the parameters here refer to the parameters of the overall model (which contains the Conv Block and the skip-connection part). The parameters of Setting 2 and Setting 3 are 51.47$\%$ and 36.03$\%$, respectively, of the parameters of Setting 1 for the traditional VM connection method, while the GFlOPs as a whole do not change much. In terms of performance, the lowest parameter of Setting 3 still maintains better segmentation performance, so in this paper, we adopt Setting 3 as the key structure of the proposed parallel Vision Mamba Layer (PVM Layer).

\textbf{Parallelization of different SSM variants.} In the Methods section, we detail the key to the influence of the parameters of Mamba, represented by SSM. However, many current studies propose improvements based on SSM to adapt it to 2D image processing. In Liu et al. \cite{liu2024vmamba}, researchers proposed 2D Selective Scan (SS2D) for visual image processing. In Runa et al. \cite{ruan2024vm}, researchers proposed VM-UNet for medical image segmentation by combining SS2D with UNet framework. In addition, based on SS2D, Wu et al. \cite{wu2024h} proposed High-order SS2D (H-SS2D) for medical image segmentation. The parameter and performance effects of SS2D and H-SS2D using parallel approach are shown in Table \ref{tab3}. From the table, it is concluded that the parameters and GFLOPs of (\emph{P}) SS2D are reduced by 81.35$\%$ and 61.97$\%$, respectively, while those of (\emph{P}) H-SS2D are reduced by 80.34$\%$ and 57.14$\%$, respectively. The above results surface that the parallel approach is effective in reducing parameters and GFLOPs not only for Mamba represented by SSM, but also for different variants of SSM at the same time.

\textbf{Plug-and-play PVM Layer.} The proposed PVM Layer can simply replace the base building blocks of any model, which include but are not limited to Convolution, Vision Transformers, Mamba, Vision Mamba and so on. Table \ref{tab4} shows the powerful plug-and-play capabilities of the PVM Layer. From the table, it can be concluded that after replacing the base building blocks or key functional modules of any model with PVM Layer, there is a significant decrease in parameters and GFLOPs, and the performance is still clearly competitive. In addition, MALUNet and EGE-UNet, as the most advanced lightweight models, after replacing the key modules in PVM Layer, the parameters and GFLOPs can still be effectively reduced, and the performance is improved. With the above results, it is shown that the powerful plug-and-play feature of PVM Layer is used to significantly reduce the parameters and GFLOPs of arbitrary models.

\section{Conclusion}
In this study, we deeply analyze the key factors of parameter influence in Mamba, and based on this, we propose a parallel Vision Mamba Layer (PVM Layer) for processing the deep features. The PVM Layer uses four Vision Mamba (VM) in parallel for processing the features, and the number of channels processed by each VM is one-fourth of the initial number of channels. This is due to the fact that the number of input channels to the SSM in Mamba has an explosive effect on the number of parameters, and the VM parameters for processing a quarter of the number of channels are 6.3$\%$ of the original VM parameters, which is an explosive reduction of 93.7$\%$. In addition, the PVM Layer can be generalized to any Mamba variant (P$x$M Layer). Based on the PVM Layer, we propose the UltraLight Vision Mamba UNet (UltraLight VM-UNet) with a parameter of only 0.049M and GFLOPs of only 0.060. The UltraLight VM-UNet parameters are 99.82$\%$ lower than those of the traditionally pure Vision Mamba UNet model (VM-UNet) and 87.84$\%$ lower than those of the lightest Vision Mamba UNet model available (LightM-UNet). In addition, we experimentally demonstrated on three publicly available skin lesion datasets that the UltraLight VM-UNet has equally strong performance competitiveness with such low parameters. Based on this paper, in the future, Mamba may become a new mainstream module for lightweight modeling.


\begin{appendix}
\section*{\Large{Appendix}}
\section{Mamba variant (SS2D) parameter impact analysis}
\begin{figure}[h]
\centering
\includegraphics[width=\linewidth]{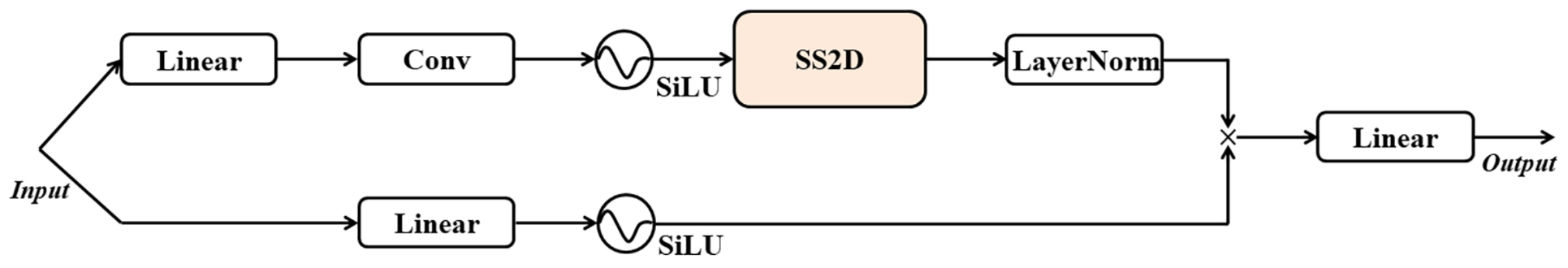}
\caption{Visual State Space (VSS) Block composition structure.}
\label{fig05}
\end{figure}

2D Selective Scan (SS2D) have been developed based on SSM which are more suitable for visual tasks, and SS2D are usually embedded in a Visual State Space (VSS) Block \cite{liu2024vmamba} as shown in Figure \ref{fig05}. The VSS Block consists of two main branches, the first one mainly consists of a linear layer and SiLU activation function \cite{elfwing2018sigmoid}. The second branch is mainly composed of linear layers, convolution, SiLU activation function, 2D Selective Scan (SS2D) and LayerNorm. Finally, the two branches merge the outputs by element-by-element multiplication.

\begin{figure}[h]
\centering
\includegraphics[width=\linewidth]{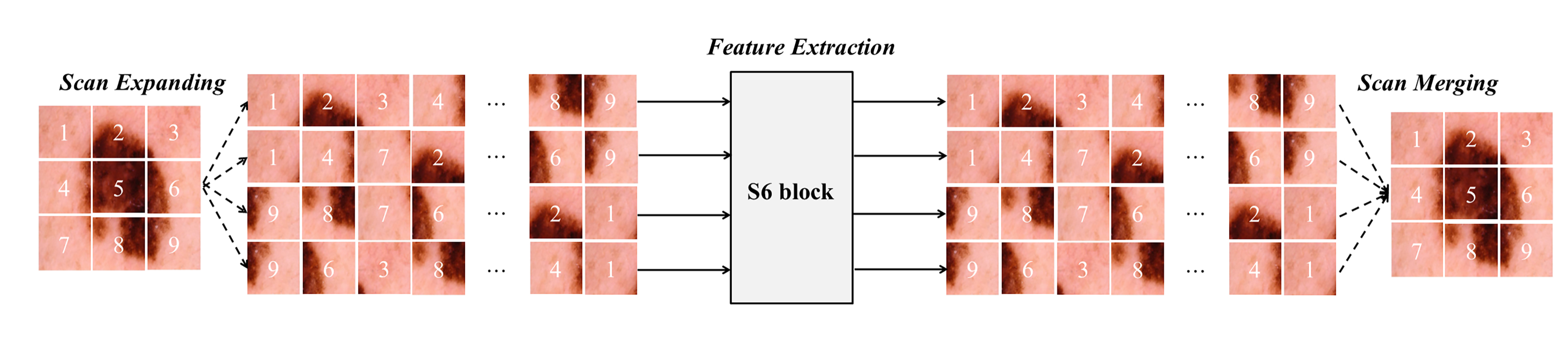}
\caption{Image interpretation of 2D Selective Scan.}
\label{fig06}
\end{figure}

The components of SS2D are shown in Figure \ref{fig06}. They include scan expansion operation, S6 block feature extraction and scan merge operation. The sequence is first expanded in four directions from top-left to bottom-right, bottom-right to top-left, top-right to bottom-left and bottom-left to top-right by scan expansion operation. Then it is input to the S6 block \cite{gu2023mamba} for feature extraction. Finally, it is merged back to the size of the original initial image by scan merge operation.

Among SS2D-based Mamba (VSS Block), the number of input channels, the size of the state dimension of the S6 block, the size of the internal convolution kernel, the projection dilation multiplier, and the rank of the projection matrix all affect the parameters. And among them, the influence of the number of input channels is explosive, and its influence mainly comes from the following aspects:

First, the $d\_inner$ of the VSS Block internal extended projection channel is determined by the product of the projection expansion multiplier and the number of input channels. This can be specifically expressed by the following equation:
\begin{equation}
d\_inner = expand *  d\_model 
\end{equation}
\noindent
where $d\_inner$ is the internal expansion projection channel, $expand$ is the projection expansion multiplier (fixed at 2 by default), and $d\_model$ is the number of input channels. We can see that $d\_inner$ will rise exponentially as the number of channels per layer in the model increases dramatically.

Second, the parameters of the input projection layer (the same input linear layer is used for both branches) and output projection layer within VSS Block will be directly related to the number of input channels. The input projection layer and output projection layer operate as follows:
\begin{equation}
in\_proj : nn.Linear(d\_model, d\_inner * 2, bias=False) 
\end{equation}
\begin{equation}
out\_proj : nn.Linear(d\_inner, d\_model, bias=False)
\end{equation}
\noindent
where then the input projection ($in\_proj$) layer parameter is $(d\_model * d\_inner * 2)$, and the output projection layer ($out\_proj$) parameter is $(d\_inner * d\_model)$. In addition, the output section has a layer normalization operation (LayerNorm) with parameter $(d\_inner + d\_inner)$. We can see that the number of input channels, $d\_model$, is the key element controlling the parameter, where the internal extended projection channel, $d\_inner$, is also controlled by $d\_model$ is controlled.

Further, the linear projection layers in the S6 block of SS2D are also key to the parameter effects. Each linear projection layer is specified as follows:
\begin{equation}
x\_proj = nn.Linear(d\_inner, (dt\_rank + d\_state * 2), bias=False) 
\end{equation}
\begin{equation}
dt\_proj = nn.Linear(dt\_rank, d\_inner, bias=True)
\end{equation}
\noindent
where $dt\_rank$ is the rank of the projection matrix $(dt\_rank = d\_model/16)$, $d\_state$ is the size of the S6 block state dimension (fixed to 16), and the parameters for each linear projection layer are $(d\_inner * (dt\_rank + d\_state*2))$. However, there are 4 linear projection layers in total, so the total parameters are $4 * (d\_inner * (dt\_rank + d\_state*2))$. In addition, $dt\_proj$ is a linear projection layer for step size with parameters $dt\_rank * d\_inner + d\_inner$, which is mainly used for linear projection for step size ($dt$). $dt\_proj$ also has 4 layers, with a total parameter of $4 * (dt\_rank * d\_inner + d\_inner)$. So, from the above, we can know that all parameters are still mainly controlled by the number of input channels $d\_model$.

In addition, the internal convolution ($nn.Conv2d(d\_inner,d\_inner,d\_conv, bias=True$)) also provides parametric influence with $d\_inner * d\_inner * d\_conv * d\_conv + d\_inner$. In this paper, $d\_conv$ is fixed to 3, so the convolution provides a parameter of $3^{2} * d\_inner^{2} + d\_inner$, which is also controlled by the $d\_model$.

Also, the $A\_logs$ of the parameter matrix controlling the attention weights of the different states of the S6 block in the SS2D module is an important influencing element. $A\_logs$ is a parameter matrix of the shape $(K*d\_inner, d\_state)$, and $K$ is a hyperparameter which is usually fixed to 4. Therefore, the parameter $A\_logs$ can be derived as $d\_inner * d\_state * 4$. In addition, the trainable vector parameter ($Ds$) of the process computed within the SS2D contains the $4 * d\_inner$ parameter, which is used to selectively integrate the SS2D state outputs with the original input signals, thereby enhancing the model's expressiveness and training stability.

In summary, assuming that the original number of input channels is 1024, keeping the other parameters unchanged, and reducing the number of channels to a quarter of the original (the number of input channels becomes 256), the original total parameters can be calculated by the above parameter formulae from 45,504,512 to 2,921,984. The parameter explosion reduces the number of channels by 93.6$\%$, which further confirms that the number of input channels has a very critical impact on the VSS Block parameters.

\section{Experimental settings}
\subsection{Datasets}
To validate that the proposed UltraLight VM-UNet also achieves competitive performance at the parameter of 0.049M, we conducted experiments on three publicly available dermatologic lesion datasets. ISIC2017 \cite{codella2018skin} and ISIC2018 \cite{codella2019skin} datasets are two large datasets published by the International Skin Imaging Collaboration (ISIC), respectively. The PH$^2$ \cite{mendoncca2013ph} dataset is a small public dataset of skin lesions, so we used PH$^2$ as an external validation to train the weights using the ISIC2017 dataset.

For the ISIC2017 dataset we acquired 2000 images as well as dermatoscope images with segmentation mask labels. Among them, the dataset was randomly divided, 1250 were used for model training, 150 images were used for model validation, and 600 images were used for model testing. The initial size of the images is 576$\times$767 pixels, and we standardize the size to 256$\times$256 pixels when inputting the model.

For the ISIC2018 dataset we acquired 2594 images as well as dermatoscope images with segmentation mask labels. Among them, the dataset was randomly divided, 1815 were used for model training, 259 images were used for model validation, and 520 images were used for model testing. The initial size of the images is 2016$\times$3024 pixels, and we standardize the size to 256$\times$256 pixels when inputting the model.

For the PH$^2$ dataset we acquired 200 images as well as dermatoscope images with segmentation mask labels. All 200 images were used for external validation. The initial size of the images was 768$\times$560 pixels and we standardized the size to 256$\times$256 pixels for inputting into the model.

\subsection{Implementation details}
The experiments were all implemented based on Python 3.8 and Pytorch 1.13.0. A single NVIDIA V100 GPU with 32GB of memory was used for all experiments. All experiments used the same data augmentation operations to more fairly determine the performance of the model, including horizontal and vertical flips, and random rotation operations. BceDice loss function was used, with AdamW \cite{loshchilov2017decoupled} as the optimizer, a training epoch of 250, a batch size of 8, and a cosine annealing learning rate scheduler with an initial learning rate of 0.001 and a minimum learning rate set to 0.00001.

\subsection{Evaluation metrics}
Dice similarity coefficient (DSC), sensitivity (SE), specificity (SP) and accuracy (ACC) are the most commonly used evaluation metrics for medical image segmentation.
DSC is used to measure the degree of similarity between the ground truth and the predicted segmentation map. SE is mainly used to measure the percentage of true positives in true positives and false negatives. SP is mainly used to measure the percentage of true negatives in true negatives and false positives. ACC is mainly used to measure the percentage of correct classification.
\begin{equation}
\mathrm{D S C}=\frac{2\mathrm{TP}}{2\mathrm{TP}+\mathrm{FP}+\mathrm{FN}}
\end{equation}
\begin{equation}
\mathrm{A C C}=\frac{\mathrm{TP}+\mathrm{TN}}{\mathrm{TP}+\mathrm{TN}+\mathrm{FP}+\mathrm{FN}}
\end{equation}
\begin{equation}
\mathrm{Sensitivity} =\frac{\mathrm{TP}}{\mathrm{TP}+\mathrm{FN}}
\end{equation}
\begin{equation}
\mathrm{Specificity} =\frac{\mathrm{TN}}{\mathrm{TN}+\mathrm{FP}}
\end{equation}
\noindent
where TP denotes true positive, TN denotes true negative, FP denotes false positive and FN denotes false negative.

\begin{figure}[h]
\centering
\includegraphics[width=\linewidth]{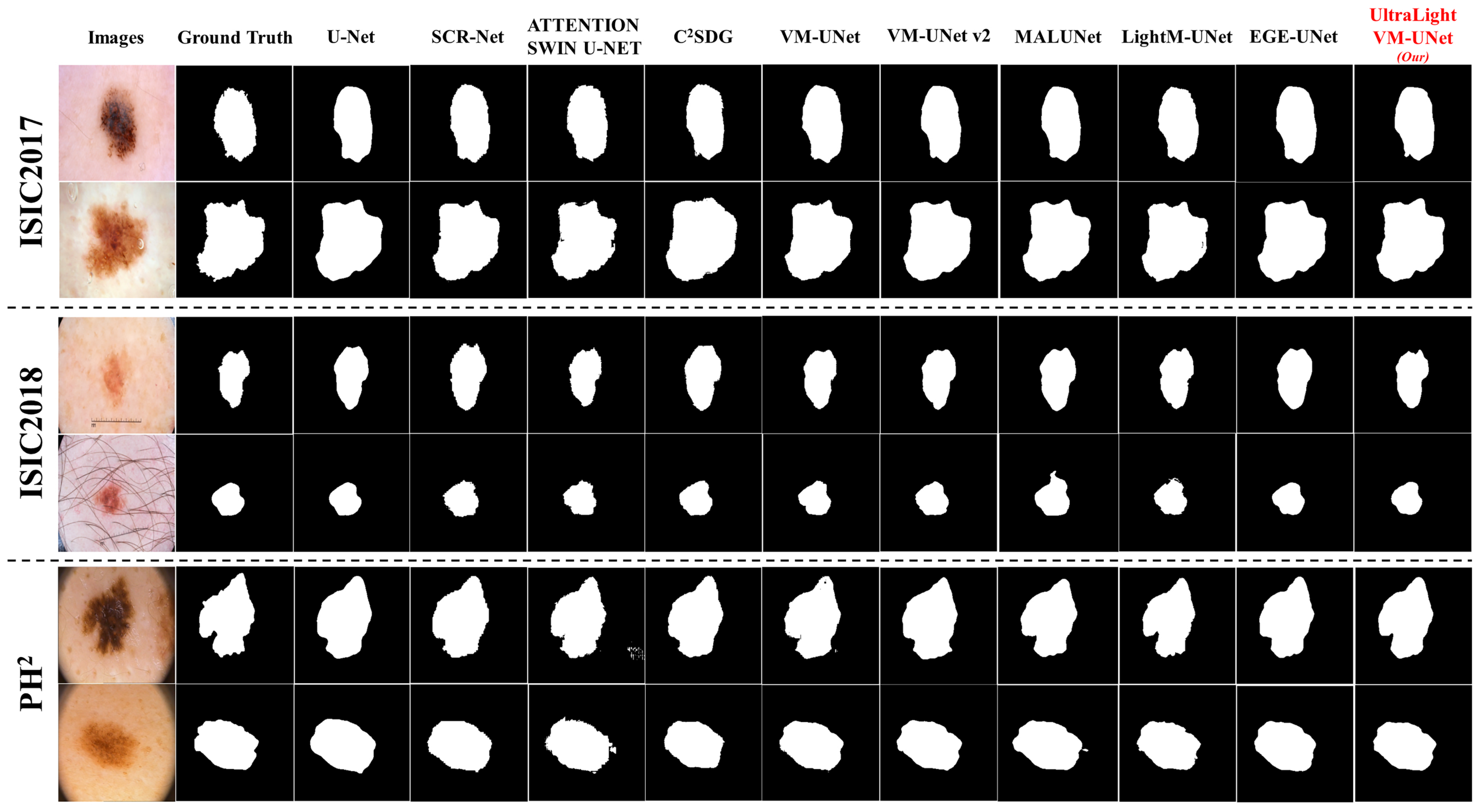}
\caption{Visualization of segmentation graphs for comparison experiments of three publicly available skin lesion segmentation datasets. }
\label{fig07}
\end{figure}

\section{Visualization results}
To more directly represent the competitive nature of UltraLight VM-UNet in terms of segmentation performance, we visualized the segmentation results (Figure \ref{fig07}). In addition, we have also visualized the segmentation results of several state-of-the-art lightweight and classical medical image segmentation models. From the visualizations, it can be concluded that the segmentation results of UltraLight VM-UNet have smooth, clear and more accurate boundaries. This shows that UltraLight VM-UNet not only outperforms the rest of the current lightweight models in terms of parameters and computational complexity, but also remains competitive in terms of performance.

\begin{table}[htbp]
  \centering
  \caption{Ablation experiments on the effect of each module in the UltraLight VM-UNet. }
  \adjustbox{width=\textwidth}{
    \begin{tabu}to 1.1\textwidth{lcc*{4}{X[c]}}\toprule
    Methods  & Params & GFLOPs & DSC & SE & SP & ACC \\\midrule
UltraLight VM-UNet (Baseline)         & 0.049M                 & 0.060  & \textbf{0.9091} & \textbf{0.9053} & 0.9790          & \textbf{0.9646} \\
Baseline\_Encoder\_Conv    & 0.080M                          & 0.071           & 0.9033          & 0.8643          & 0.9880          & 0.9638          \\
Baseline\_Decoder\_Conv    & 0.080M                          & 0.064           & 0.8958          & 0.8512          & \textbf{0.9881} & 0.9612          \\
Baseline\_(En+De)\_Conv    & 0.123M                          & 0.075           & 0.9065          & 0.8784          & 0.9855          & 0.9645 \\
Baseline\_SCAB not applicable    & \textbf{0.033M}                          & \textbf{0.058}           & 0.9029          & 0.8767          & 0.9841          & 0.9631 \\\bottomrule
    \end{tabu}%
    } \vspace{-5pt}
  \label{tab5}%
\end{table}%

\section{More analysis}
\subsection{Impact of components in the UltraLight VM-UNet}
A series of ablation experiments were performed to verify the impact of each module in the UltraLight VM-UNet. As shown in Table \ref{tab5}, we replace the PVM Layer of the encoder, decoder, respectively, with a standard convolution with convolution kernel 3. In addition, we also replace the PVM Layer of the encoder and decoder at the same time with a standard convolution. Also, $Baseline\_SCAB \enspace not \enspace applicable$ indicates that the skip-connection part of the UltraLight VM-UNet does not use the SAB and CAB modules. From the table, we can conclude that in replacing the PVM Layer of the encoder and decoder separately, the parameters increased by 63.26$\%$ and the GFLOPs increased in both, while the performance decreased in both. In particular, after replacing the PVM Layer of the encoder and decoder simultaneously, the parameters increase by 151$\%$ and the GFOLPs are increased by 25$\%$. In summary, it is shown that after replacing the PVM Layer there is a decrease in all performance aspects and an increase in both parameters and GFLOPs. This again proves the crucial role of PVM Layer. What's more, although the parameters and GFOLPs were further reduced with the removal of the SAB and CAB modules. However, the performance also exhibits some degradation, which is due to the ability of the SAB and CAB modules to combine multi-scale feature information for learning, which improves the sensitivity to lesions and accelerates the model convergence \cite{ruan2022malunet}. Therefore, in order to balance the performance and computational complexity relationship, UltraLight VM-UNet employs the SAB and CAB modules as a bridge for skip connections to further improve segmentation performance. Even so, the performance, parameters and GFLOPs of UltraLight VM-UNet are still in the forefront compared to the rest of the state-of-the-art lightweight models available currently.

\begin{table}[htbp]
  \centering
  \caption{Ablation experiments on the effect of combining different channel numbers in the UltraLight VM-UNet. D$^{\rm P}$ denotes the percentage of parameter reduction with the parallel approach. 74.80$\%$ denotes the theoretical percentage reduction of localized module parameters derived from the Mamba analysis in the methods section. {\color{02pink}$\maltese$} indicates no parallel connection method and {\color{00red}$\bigstar$} indicates a quadruple parallel connection method. }
  \adjustbox{width=\textwidth}{
    \begin{tabu}to 1.1\textwidth{lccc*{4}{X[c]}}\toprule
    Methods & Params  & D$^{\rm P}$(74.80$\%$) & GFLOPs & DSC & SE & SP & ACC \\\midrule
{\color{02pink}$\maltese$} $[8,16,24,32,48,64]$        & 0.136M    & \multirow{2}{*}{63.98$\%$}$_{\textcolor{red}{\text{(-10.82$\%$)}}}$    & 0.060  & 0.9069 & 0.8861 & 0.9834          & 0.9644 \\
{\color{00red}$\bigstar$} $[8,16,24,32,48,64]$        & 0.049M    &        & 0.060  & 0.9091 & 0.9053 & 0.9790       & 0.9646 \\\midrule
{\color{02pink}$\maltese$} $[8,16,32,64,128,256]$        & 0.909M    & \multirow{2}{*}{75.47$\%$}$_{\cb{\text{(+0.67$\%$)}}}$    & 0.074  & 0.9018 & 0.8749 & 0.9840          & 0.9627 \\
{\color{00red}$\bigstar$} $[8,16,32,64,128,256]$        & 0.223M    &        & 0.074  & 0.9079 & 0.8965 & 0.9809          & 0.9644 \\\midrule
{\color{02pink}$\maltese$} $[16,32,64,128,256,512]$        & 3.479M    & \multirow{2}{*}{75.63$\%$}$_{\cb{\text{(+0.83$\%$)}}}$    & 0.259  & 0.9049 & 0.8981 & 0.9789          & 0.9630 \\
{\color{00red}$\bigstar$} $[16,32,64,128,256,512]$        & 0.848M    &        & 0.259  & 0.9056 & 0.8906 & 0.9815          & 0.9637 \\\midrule
{\color{02pink}$\maltese$} $[32,64,128,256,512,1024]$        & 13.607M    & \multirow{2}{*}{75.71$\%$}$_{\cb{\text{(+0.91$\%$)}}}$    & 0.970  & 0.9053 & 0.8878 & 0.9821          & 0.9637 \\
{\color{00red}$\bigstar$} $[32,64,128,256,512,1024]$        & 3.305M    &        & 0.970  & 0.9012 & 0.8812 & 0.9819          & 0.9622 \\\bottomrule
    \end{tabu}%
    } \vspace{-5pt}
  \label{tab6}%
\end{table}%

\subsection{Impact of different channel numbers in UltraLight VM-UNet}
In addition, in order to verify the impact of different channel numbers in the UltraLight VM-UNet, we conducted a series of ablation experiments. As shown in Table \ref{tab6}, we set up four different combinations of channel numbers, including $[8,16,24,32,48,64]$, $[8,16,32,64,128,256]$, $[16,32,64,128,256,512]$ and $[32,64,128,256,512,1024]$, respectively. In addition, we conducted experiments both in the parallel-free manner (labeled by {\color{02pink}$\maltese$}) and in the parallel manner (labeled by {\color{00red}$\bigstar$}). From the table, it can be concluded that the parameters and GFLOPs at $[8,16,24,32,48,64]$ are the lowest, while the overall difference in performance is not very large. However, as the number of channels increases, both parameters and GFLOPs show a significant rise. The parameter increases from 0.136M to 13.607M for the group without parallelism and from 0.049M to 3.305M for the group with parallelism. In particular, the average decrease in parameters for the same number of channel combinations using the parallel approach over the no-parallel approach is 72.70$\%$, which is extremely close to the percentage of decrease in the four-parallel approach over the no-parallel approach analyzed in the Methods section (74.80$\%$). In addition, at $[8,16,24,32,48,64]$, the parameters are more susceptible to the rest of the modules, so the decrease is not as large as for the rest of the channel number combinations. However, its overall parameters and GFLOPs reach an impressive 0.049M and 0.060, and show strong competitive segmentation performance.

\end{appendix}

\end{document}